\documentclass[english,aps,twocolumn,pra]{revtex4}

\usepackage[T1]{fontenc}

\usepackage{graphicx}
\usepackage{epstopdf}
\usepackage{amssymb}
\usepackage{amsmath}
\usepackage[urlcolor=blue,hyperindex,colorlinks,bookmarks=true,linkcolor=black,citecolor=black]{hyperref}
\usepackage[normalem]{ulem}
\usepackage[abs]{overpic}
\usepackage{color}
\usepackage{rotating}
\usepackage{subfigure}

\usepackage[english]{babel}
\usepackage{blindtext,tikz}
\usetikzlibrary{calc}

\newcommand{\be}{\begin{equation}}
\newcommand{\ee}{\end{equation}}
\newcommand{\bea}{\begin{align}}
\newcommand{\eea}{\end{align}}

\newcommand{\ket}[1]{\left|#1\right\rangle}
\newcommand{\bra}[1]{\left\langle#1\right|}

\newcommand{\abs}[1]{\lvert#1\rvert}

\newcommand{\al}{\alpha}
\newcommand{\ww}{\omega}
\newcommand{\eps}{\epsilon}

\begin{document}

\title{Coherent Feedback Improved Qubit Initialization in the Dispersive Regime}

\author{Luke C.G. Govia}
\email[Electronic address: ]{lcggovia@lusi.uni-sb.de}
\affiliation{Theoretical Physics, Universit\"{a}t des Saarlandes, Campus, 66123 Saarbr\"{u}cken, Germany}
\author{Frank K. Wilhelm}
\affiliation{Theoretical Physics, Universit\"{a}t des Saarlandes, Campus, 66123 Saarbr\"{u}cken, Germany}

\begin{abstract}
Readout of the state of a superconducting qubit by homodyne detection of the output signal from a dispersively coupled microwave resonator is a common technique in circuit quantum electrodynamics, and is often claimed to be quantum non-demolition (QND) up to the same order of approximation as in the dispersive approximation. However, in this work we show that only in the limit of infinite measurement time is this protocol QND, as the formation of a dressed coherent state in the qubit-cavity system applies an effective rotation to the qubit state. We show how this rotation can be corrected by a coherent operation, leading to improved qubit initialization by measurement and coherent feedback.
\end{abstract}

\maketitle

For most quantum information and computing protocols measurement is a necessary component, either to extract the answer to a computation, or as an operation in the protocol, such as for entanglement generation or gate operations. In addition, many protocols benefit from so called quantum non-demolition (QND) measurement, where the Hamiltonian describing the measurement operator commutes with the self-Hamiltonian of the system \cite{Braginsky:1992kq}. As a result, perfect QND measurement maximally dephases the system in its eigenbasis, and the system state is projected onto an eigenstate when the measurement result is observed. Alternatively, one can think of a QND measurement as having only the minimal (required by quantum mechanics) back action on the system it measures.

In the field of circuit quantum electrodynamics (cQED), the state of a superconducting qubit is typically measured in its eigenbasis by homodyne detection of the phase of the output signal through a cavity dispersively coupled to the qubit \cite{Blais2004,Johansson:2006vn}. Due to the small signal strength exiting the cavity, it is necessary to amplify the signal using a low noise (near quantum limited) parametric amplifier based on the nonlinearity induced by a Josephson junction \cite{Siddiqi:2004fj,Castellanos07,Bergeal:2010ty,Bishop:2010qf,Reed:2010kq,Mutus:2014yg,OBrien:2014yg,White2015}. In recent years this measurement scheme has been a great success, with highlights that include the observation of qubit quantum jumps \cite{Vijay2011}, heralded initialization via measurement \cite{Johnson12,Riste12}, entanglement generation between qubits \cite{Riste2013,Chow:2014fk}, quantum teleportation \cite{Steffen2013}, and readout fidelity greater than 99\% \cite{Jeffrey:2014zr}.

Under the dispersive approximation, this readout scheme has been reported in the literature to be QND \cite{Blais2004}, as in the dispersive frame the qubit-cavity coupling is diagonal and commutes with the system self-Hamiltonian. However, it was recently shown \cite{longpaper} that for a semi-classically driven cavity, the joint system of the qubit-cavity in the lab frame is an entangled state known as the dressed coherent state. To lowest order, this entanglement results in a rotation of the qubit state that depends on the coherent state amplitude in the cavity. As a result, dispersive measurement as previously proposed is not perfectly QND, even up to the same order of approximation as the dispersive approximation, except in the limit of infinite measurement time. This poses problems for schemes that require perfect QNDness, such as those performing heralded initialization or entanglement generation \cite{Johnson12,Riste12,Hutchison:2009qy}.

In this letter we examine the dispersive qubit readout protocol and account for the effects of the formation of dressed coherent states during the protocol. In particular, we describe the effective coherent qubit rotation that depends on both the amplitude and phase of the applied cavity drive. This rotation is equivalent to a change of the measurement basis, and, as it is coherent, it can be corrected for by unitary feedback. This opens up the possibility for true QND measurement by introducing coherent feedback.

We consider a system consisting of a single qubit coupled to a microwave resonator (cavity), as described by the familiar Jaynes-Cummings Hamiltonian \cite{Jaynes63}
\begin{align}
\hat{H} = \ww_{\rm c}\hat{a}^{\dagger}\hat{a} - \frac{\ww_{\rm q}}{2}\hat{\sigma}_z + g\left(\hat{\sigma}^{-}\hat{a}^{\dagger}+\hat{\sigma}^{+}\hat{a}\right),
\label{eqn:JC}
\end{align}
where $\hat{a}$ and $\hat{a}^{\dagger}$ are the usual bosonic annihilation and creation operators for the cavity, $\hat{\sigma}_z$ is the Pauli matrix whose eigenstates are the qubit logical states, $\hat{\sigma}^{\pm}$ are the qubit raising and lowering operators, $\ww_{\rm c/q}$ are the cavity and qubit frequencies, $g$ is the Jaynes-Cummings coupling strength, and we set $\hbar = 1$ from here on. This Hamiltonian describes evolution in the lab frame, by which we mean we have not described any of the system's evolution by a (possibly time dependent) rotation of Hilbert space.

After the dispersive frame transformation, in the limit $\lambda = g/\Delta <1$, where $\Delta = \ww_{\rm q} - \ww_{\rm c}$, the Jaynes-Cummings Hamiltonian reduces to the dispersive Hamiltonian
\be
\hat{H}_{\rm D} = \ww_{\rm c}\hat{a}^{\dagger}\hat{a} - \frac{\ww_{\rm q}+\chi}{2}\hat{\sigma}_z -\chi\hat{\sigma}_z \hat{a}^{\dagger}\hat{a},
\label{eqn:Disp}
\ee
where $\chi = g^2/\Delta$, and we have kept terms only up to second order in $\lambda$. The dispersive frame transformation, followed by the discarding of terms beyond second order in $\lambda$ is commonly called the dispersive approximation. Under the dispersive approximation, the system eigenstates in the lab frame of equation (\ref{eqn:JC}) are \cite{Haroche06}
\begin{align}
&\overline{\ket{g,n}} = \cos\left({\lambda\sqrt{n}}\right)\ket{g,n} - \sin\left({\lambda\sqrt{n}}\right)\ket{e,n-1}, \label{eqn:aGeig} \\
&\overline{\ket{e,n-1}} = \cos\left({\lambda\sqrt{n}}\right)\ket{e,n-1} + \sin\left({\lambda\sqrt{n}}\right)\ket{g,n}, \label{eqn:aEeig} 
\end{align}
which are referred to as the dressed eigenstates, and it is worth pointing out that $\overline{\ket{g,0}} = \ket{g,0}$, i.e. the dressed and undressed ground states are the same as $\ket{g,0}$ is dark.

The last required ingredient for dispersive qubit-state readout is a classical cavity drive, described by the Hamiltonian
\be
\hat{H}_{\rm d}(t) = 2\cos(\ww_{\rm d}t)\left(\eps\hat{a}+\eps^*\hat{a}^{\dagger}\right),
\label{eqn:CavDri}
\ee
in the lab frame. Under the dispersive approximation this Hamiltonian is unaffected to lowest order in $\lambda$, and the leading order correction term is both damped by the small parameter $\lambda$ and oscillates quickly provided $\ww_{\rm d} \neq \ww_{\rm q}$. Now, if we consider photons from the applied drive that interact with the qubit-cavity system, when they exit the cavity they will carry qubit information with them which can be used to read out the state of the qubit. In particular, by setting $\ww_{\rm d} = \ww_{\rm c}$ the qubit state information is contained only in the phase of the signal exiting the cavity, as described in Ref.~\cite{Blais2004}. These statements will be made more concrete shortly.

As was shown in \cite{longpaper}, if the system starts in either initial state $\overline{\ket{g/e,0}}$ (see appendix \ref{app:UndressedE} for the bare excited state as the initial state), then after applying a cavity drive of the form of equation (\ref{eqn:CavDri}) for a time $t_{\rm d}$ the state of the qubit-cavity system in the lab frame will be the dressed coherent state $\overline{\ket{g/e,\alpha_{g/e}(t_{\rm d})}}$, defined by
\begin{align}
\overline{\ket{g/e,\alpha_{g/e}(t_{\rm d})}} = e^{-\frac{\abs{\alpha_{g/e}(t_{\rm d})}^2}{2}}\sum_{n}\frac{\alpha_{g/e}(t_{\rm d})^n}{\sqrt{n!}}\overline{\ket{g/e,n}},
\end{align}
where $\alpha_{g/e}(t_{\rm d})$ are given by
\begin{align}
\nonumber\al_{\rm g}(t_{\rm d}) &= \frac{\eps^*}{\chi}\left(e^{-i\chi t_{\rm d}}-1\right)e^{-i(\ww_{\rm c}-\chi)t_{\rm d}} \\ \nonumber&=  -\frac{2i\eps^*}{\chi}\sin\left(\frac{\chi}{2}t_{\rm d}\right)e^{-i(\ww_{\rm c}-\frac{\chi}{2})t_{\rm d}},\\
\nonumber\al_{\rm e}(t_{\rm d}) &= \frac{-\eps^*}{\chi}\left(e^{i\chi t_{\rm d}}-1\right)e^{-i(\ww_{\rm c}+\chi)t_{\rm d}} \\&=  -\frac{2i\eps^*}{\chi}\sin\left(\frac{\chi}{2}t_{\rm d}\right)e^{-i(\ww_{\rm c}+\frac{\chi}{2})t_{\rm d}},
\label{eqn:Alphas} 
\end{align}
for $\ww_{\rm d} = \ww_{\rm c}$ as used for dispersive readout. The phase factors $e^{-i(\ww_{\rm c}\pm\chi)t_{\rm d}}$ are due to the cavity self-Hamiltonian as well as the dispersive interaction. The dressed coherent state is entangled, and correctly accounts for the correlations created between the qubit and the cavity during the applied classical drive.

To first order in $\lambda$, both dressed coherent states can be approximated by (see appendix \ref{app:DressApp} for further details)
\begin{align}
&\overline{\ket{g,\alpha_{g}(t_{\rm d})}} = \left(\ket{g}-\lambda\al_{g}(t_{\rm d})\ket{e}\right)\ket{\alpha_{g}(t_{\rm d})}/\sqrt{\mathcal{N}}+ \mathcal{O}(\lambda^2), \\
&\overline{\ket{e,\alpha_{e}(t_{\rm d})}} = \left(\ket{e}+\lambda\al^*_{e}(t_{\rm d})\ket{g}\right)\ket{\alpha_{e}(t_{\rm d})}/\sqrt{\mathcal{N}}+ \mathcal{O}(\lambda^2), 
\end{align}
where $\mathcal{N} = 1 +\lambda^2\abs{\al_{g/e}(t_{\rm d})}^2$, and even for $\lambda \ll 1$ we can keep the term proportional to $\lambda\abs{\alpha_{g/e}(t_{\rm d})}$ as $\abs{\alpha_{g/e}(t_{\rm d})}$ can be large. This approximation gives a good intuitive picture of the effect of an applied cavity drive on a qubit-cavity system described in the lab frame. The cavity is driven to a coherent state (as expected), while the qubit state is rotated a small amount. This rotation depends on the coherent state amplitude $\alpha_{g/e}(t_{\rm d})$, and therefore on the amplitude, phase, and duration of the applied cavity drive.

To connect to dispersive readout \cite{Blais2004}, we introduce a cavity decay mechanism via the cavity-environment coupling operator $\hat{a} + \hat{a}^\dagger$, described for a bare cavity by the quality factor $Q_{\rm F}$. For an approximately Ohmic environment around the cavity frequency, such as for an open transmission line, the decay rate is defined in terms of the quality factor by $\kappa(\ww_{\rm c}) = \ww_{\rm c}/Q_{\rm F}$. For a coupled qubit-cavity system, following the dressed decoherence model of \cite{Beaudoin11,Sete:2014fk}, in addition to cavity decay there will also be cavity-mediated qubit decay (indirect Purcell decay \cite{Beaudoin11}). To lowest order in $\lambda$ (as shown in \cite{Beaudoin11,Sete:2014fk}), for an approximately Ohmic environment around the qubit frequency, this occurs at a rate $\gamma_{\rm P} = \lambda^2(\ww_{\rm q}+\chi)/Q_{\rm F}$ regardless of the cavity photon number (photon number effects become relevant at higher orders of $\lambda$). In experiment, Purcell decay can be almost completely removed by appropriate filtering of the cavity output at the qubit frequency, a technique known as Purcell filtering \cite{Reed10,Jeffrey:2014zr}.

In the eigenbasis of the Jaynes-Cummings Hamiltonian, the two decay mechanisms described above amount to the following. ``Cavity decay'' is the dressed eigenstate transition $\overline{\ket{g/e,n+1}} \rightarrow \overline{\ket{g/e,n}}$, which effectively preserves the qubit state, while Purcell decay is the dressed eigenstate transition $\overline{\ket{e,n}} \rightarrow \overline{\ket{g,n}}$, which effectively preserves the photon number in the cavity. All other transitions have zero matrix elements with the cavity-environment coupling operator and are therefore forbidden.

Initially, let us assume that we can neglect Purcell decay, as would be the case if a suitable Purcell filter is connected to the cavity, as has been achieved in state of the art dispersive readout \cite{Reed10,Jeffrey:2014zr}. In addition, we assume that the cavity only begins decaying after the state $\overline{\ket{g/e,\alpha_{g/e}(t_{\rm d})}}$ has been created, and we assume a temperature of zero. For this simplified case, given that only transitions of the form $\overline{\ket{g/e,n+1}} \rightarrow \overline{\ket{g/e,n}}$ are allowed, we see that the dressed coherent state will decay similarly to a coherent state in a bare cavity, such that after a time $\tau$ of decay the qubit-cavity state will be
\newpage
\begin{align}
&\nonumber\overline{\ket{g,\alpha_{g}(t_{\rm d})e^{-\frac{\kappa}{2}\tau}e^{-i(\ww_{\rm c}-\chi)\tau}}} \\ \nonumber&= \left(\ket{g}-\lambda\al_{g}(t_{\rm d})e^{-\frac{\kappa}{2}\tau}e^{-i(\ww_{\rm c}-\chi)\tau}\ket{e}\right)\\ &\otimes\ket{\alpha_{g}(t_{\rm d})e^{-\frac{\kappa}{2}\tau}e^{-i(\ww_{\rm c}-\chi)\tau}}/\sqrt{\mathcal{N(\tau)}} + \mathcal{O}(\lambda^2),
\label{eqn:apDg} \\
&\nonumber\overline{\ket{e,\alpha_{e}(t_{\rm d})e^{-\frac{\kappa}{2}\tau}e^{-i(\ww_{\rm c}+\chi)\tau}}} \\ \nonumber&= \left(\ket{e}+\lambda\al^*_{e}(t_{\rm d})e^{-\frac{\kappa}{2}\tau}e^{i(\ww_{\rm c}+\chi)\tau}\ket{g}\right)\\ &\otimes\ket{\alpha_{e}(t_{\rm d})e^{-\frac{\kappa}{2}\tau}e^{-i(\ww_{\rm c}+\chi)\tau}}/\sqrt{\mathcal{N(\tau)}} + \mathcal{O}(\lambda^2),
\label{eqn:apDe}
\end{align}
which is just a dressed coherent state with a damped amplitude $\abs{\alpha_{g/e}(t_{\rm d})}e^{-\frac{\kappa}{2}\tau}$.

Following equations (\ref{eqn:apDg}) and (\ref{eqn:apDe}), after a cavity decay of time $\tau$ the initial qubit states $\ket{g/e}$ have been mapped approximately to
\begin{flalign}
&\nonumber \ket{g} \rightarrow \\&\left(\ket{g}-\lambda\abs{\al}e^{-i(\varphi_\epsilon + \pi/2)}e^{-i\ww_gt_{\rm d}}e^{-\frac{\kappa}{2}\tau}e^{-i\ww'_g\tau}\ket{e}\right)/\sqrt{\mathcal{N}(\tau)}, \label{eqn:dispRg} \\
&\nonumber\ket{e} \rightarrow \\&\left(\ket{e}+\lambda\abs{\al}e^{i(\varphi_\epsilon + \pi/2)}e^{i\ww_et_{\rm d}}e^{-\frac{\kappa}{2}\tau}e^{i\ww'_e\tau}\ket{g}\right)/\sqrt{\mathcal{N}(\tau)}, \label{eqn:dispRe}
\end{flalign}
where we have used the fact that $\al_{g/e}(t_{\rm d})$ differ in phase only to define $\al = \abs{\al_{g/e}(t_{\rm d})}$ and the phase factor $-(\varphi_\epsilon + \pi/2)$ of $-i\epsilon^*$, as well as the shifted cavity frequencies $\ww_{g/e} = \ww_{\rm c}\mp\frac{\chi}{2}$ and $\ww'_{g/e} = \ww_{\rm c}\mp\chi$. The measured phases of the output signal jump sharply when the drive pulse is turned off, with the coherent states rotating around phase space at frequencies $\ww_{g/e}$ for times $t\leq t_{\rm d}$ and at frequencies $\ww'_{g/e}$ for times $\tau = t - t_{\rm d} > 0$.

The qubit-state maps of equations (\ref{eqn:dispRg}) and (\ref{eqn:dispRe}) can alternatively be understood as qubit-cavity interactions during the cavity drive changing the basis of qubit measurement, with measurement of the cavity frequency $\ww_g$/$\ww'_g$ for a total time $t_{\rm d}+ \tau$ corresponding to the qubit state $\ket{g}-\lambda\abs{\al}e^{-\frac{\kappa}{2}\tau}e^{-i(\varphi_\epsilon + \pi/2 + \ww_gt_{\rm d} + \ww'_g\tau)}\ket{e}$ and a similar result for the measurement of $\ww_e$/$\ww'_e$. Only in the $\tau \rightarrow \infty$ limit does the basis of measurement become $\left\{\ket{g},\ket{e}\right\}$ and the measurement QND. 

The key observation in this work is that the change of basis of measurement is a coherent rotation error applied to the qubit output state. As this error in the qubit state is coherent, it can be actively corrected for by a single qubit rotation that applies the inverse of the unitary map of equation (\ref{eqn:dispRg}) or (\ref{eqn:dispRe}). This conditional rotation to correct the output state is defined by the unitary operators
\begin{flalign}
&\hat{U}_{g}\left(t_{\rm d},\tau,\abs{\alpha}\right) = \exp\left\{i\left(\cos(\Sigma_g)\hat{\sigma}_y-\sin(\Sigma_g)\hat{\sigma}_x\right)\theta\right\}, \label{eqn:Ug}\\
&\hat{U}_{e}\left(t_{\rm d},\tau,\abs{\alpha}\right) =\exp\left\{i\left(\cos(\Sigma_e)\hat{\sigma}_y-\sin(\Sigma_e)\hat{\sigma}_x\right)\theta\right\}, \label{eqn:Ue}
\end{flalign}
where $\Sigma_g = \ww_gt_{\rm d}+\ww'_g\tau+\varphi_\epsilon + \pi/2$, $\Sigma_e =\varphi_\epsilon + \pi/2+\ww_et_{\rm d}+\ww'_e\tau$, and $ \tan(\theta) = \lambda\abs{\al}e^{-\frac{\kappa}{2}\tau}$.

While this erroneous qubit rotation scales as $\lambda$ and $e^{-\frac{\kappa}{2}\tau}$ and is therefore small, it is an effect that will propagate throughout a computation, affecting the fidelity of all subsequent gates and measurements. Therefore, the observation that it can be easily corrected for is a useful one, especially in architectures where qubit measurements are used as a means of initialization at the beginning of a computation \cite{Johnson12,Riste12}, or used during the computation to stabilize error correction codes \cite{Fowler:2012sf}, and in light of the fact that dispersive qubit readout fidelity approaches ever higher values \cite{Jeffrey:2014zr}.

\begin{figure*}
\subfigure{
\includegraphics[width = \columnwidth]{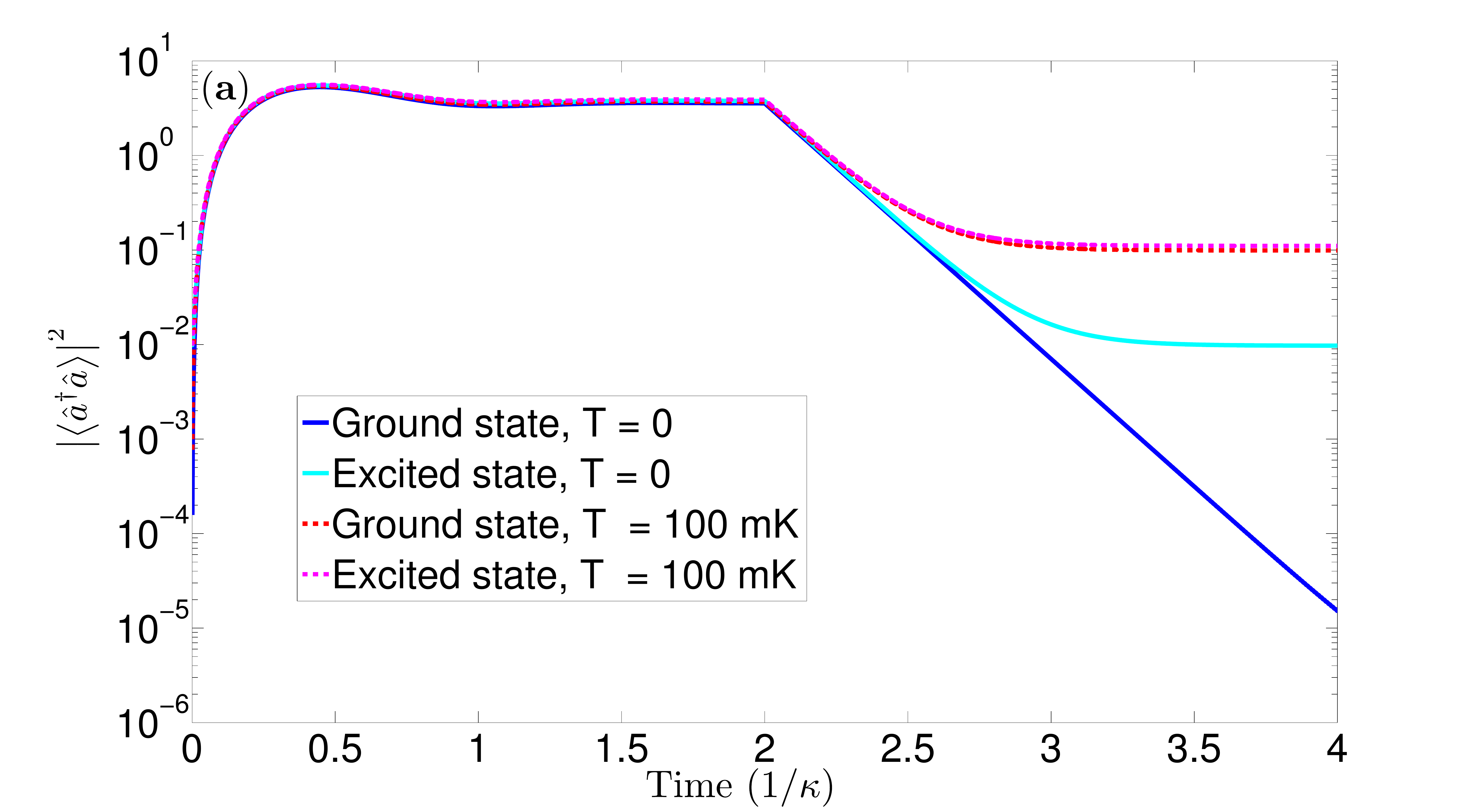}
\label{fig:CavOc}}
\subfigure{
\includegraphics[width = \columnwidth]{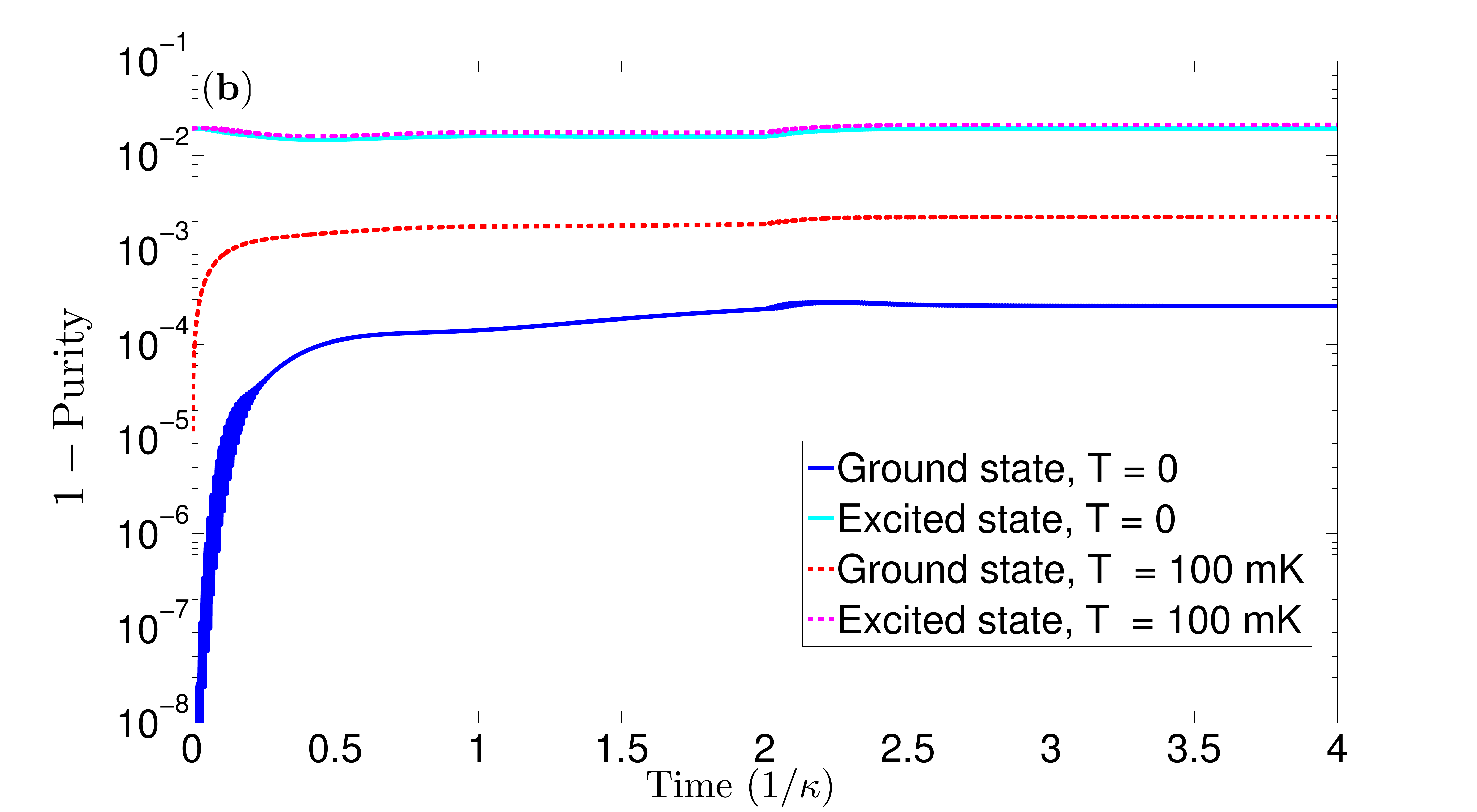}
\label{fig:Pur}}
\subfigure{
\includegraphics[width = \columnwidth]{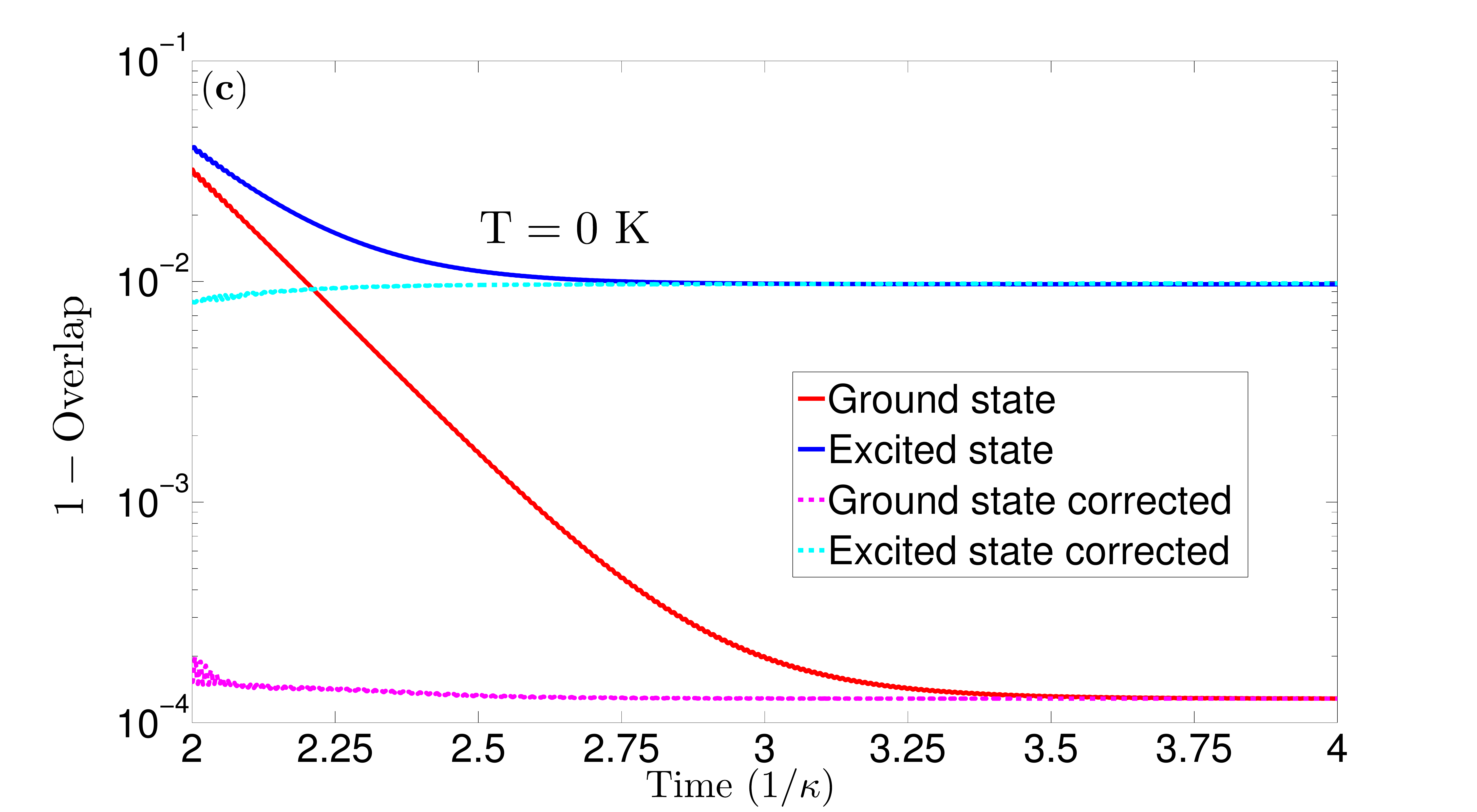}
\label{fig:Cold}}
\subfigure{
\includegraphics[width = \columnwidth]{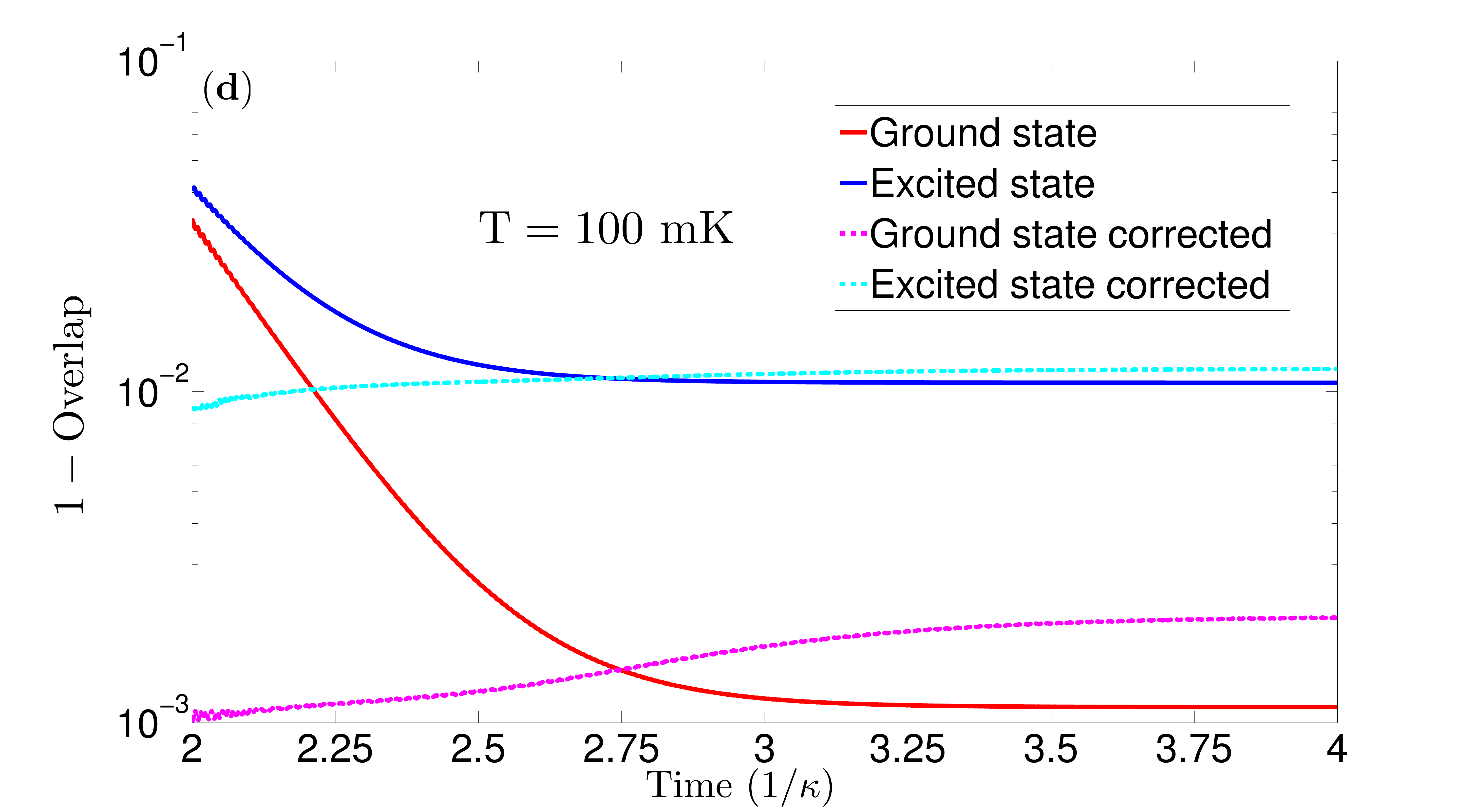}
\label{fig:Hot}}
\caption{{\bf (a)} Cavity occupation, and {\bf (b)} $1 - P(t)$ for T = 0 and T = 100 mK for both initial states. $1 - \mathcal{F}_{\nu}(\tau)$ and $1 - \mathcal{F}_{\nu}^{\rm C}(\tau)$ are shown in {\bf (c)} for T = 0 and {\bf (d)} for T = 100 mK. A drive strength of $\abs{\eps}/2\pi = 0.04$ GHz, a cavity decay rate of $1/\kappa = 100$ ns, and $\abs{\lambda} = 0.1$ were used for these simulations.}
\label{fig:Res}
\end{figure*}

In order to relax our previous assumption of an initially closed cavity, and to consider effects beyond first order in $\lambda$, we numerically simulate the cavity drive and decay process with the Purcell filtered master equation (see appendix \ref{app:NME} for further details, and appendix \ref{app:Pdecay} for simulations with Purcell decay)
\begin{align}
&\dot{\rho}(t) = -i\left[\hat{H}_{\rm T}(t),\rho(t)\right] \label{eqn:MEnoP} \\ \nonumber&+ \kappa\left(\left(1+ n_{\rm th}(\ww_{\rm c},{\rm T})\right)\mathcal{D}\Big[\hat{a}_{\mathcal{C}}\Big] + n_{\rm th}(\ww_{\rm c},{\rm T})\mathcal{D}\left[\hat{a}^{\dagger}_{\mathcal{C}}\right]\right)\rho(t),
\end{align}
where the operator $\hat{a}_{\mathcal{C}}$ of equation (\ref{eqn:OpNoPur}) describes only cavity decay, $ n_{\rm th}(\ww,{\rm T})$ is the Bose distribution at frequency $\ww$ and temperature T, and $\mathcal{D}(x)\rho$ is the dissipator defined by
\be
\mathcal{D}(x)\rho = x\rho x^\dagger - \frac{1}{2}\left\{x^\dagger x,\rho\right\}.
\ee
Here the total system Hamiltonian describes the full Jaynes-Cummings interaction between the qubit and the cavity as well as the classical cavity drive, and is given by
\begin{align}
\nonumber\hat{H}_{\rm T}(t) &= \ww_{\rm c}\hat{a}^{\dagger}\hat{a} - \frac{\ww_{\rm q}}{2}\hat{\sigma}_z + g\left(\hat{\sigma}^{-}\hat{a}^{\dagger}+\hat{\sigma}^{+}\hat{a}\right) \\&+ \left(\eps e^{i\ww_{\rm d} t}\hat{a}+\eps^*e^{-i\ww_{\rm d} t}\hat{a}^{\dagger}\right)\Theta(t-t_{\rm d}),
\end{align}
where $\Theta(x)$ is the Heaviside step function. We simulate the evolution for the initial states $\ket{g,0}$ and $\overline{\ket{e,0}}$, with the temperature set at either T = 0 or T = 100 mK. The results are shown in FIG.~\ref{fig:Res}.

FIG.~\ref{fig:CavOc} shows the cavity occupation during the readout protocol. As expected, after an initial ring-up phase, once the drive is turned off the cavity occupation decays. Decay stops once the steady state is reached, which is $\ket{g,0}$ or $\overline{\ket{e,0}}$ for T = 0 and thermally broadened versions of these states for T = 100 mK. FIG.~\ref{fig:Pur} shows $1 - P(t)$, where $P(t)$ is the purity of the qubit state, defined for a reduced qubit state $\rho(t)$ by $P(t) = {\rm Tr}[\rho(t)^2]$. Unit purity indicates a pure state. As can be seen, for T = 0 the states remain very close to a pure state at all times, verifying the analytic results of equations (\ref{eqn:apDg}) and (\ref{eqn:apDe}). Even for T = 100 mK the states remain $>90\%$ pure for either initial state.

As the states remain mostly pure during the protocol, it is possible to correct the qubit state error by the unitaries of equations (\ref{eqn:Ug}) and (\ref{eqn:Ue}), as described previously. To quantify this correction we use the overlap between the desired state ($\ket{g}$ or $\ket{e}$) and the simulated reduced qubit state $\rho(t)$. We measure the overlap before correction
\be
\mathcal{F}_{\nu}(\tau) = {\rm Tr}\left[\ket{\nu}\bra{\nu}\rho(t)\right],
\ee
where the subscript $\nu\in\{g,e\}$ indicates whether we started in $\ket{g,0}$ or $\overline{\ket{e,0}}$, and the overlap after correction
\be
\mathcal{F}_{\nu}^{\rm C}(\tau) = {\rm Tr}\left[\ket{\nu}\bra{\nu}\hat{U}_{\nu}\rho(t)\hat{U}^{\dagger}_{\nu}\right].
\ee
FIG.~\ref{fig:Cold} shows the overlap error for both the uncorrected and the corrected state for T = 0, and as can be seen $\mathcal{F}_{\nu}^{\rm C}(\tau) \ge \mathcal{F}_{\nu}(\tau)$ for all time (to within numerical precision of the simulations). For T = 100 mk, as shown in FIG.~\ref{fig:Hot}, this is not the case, as within roughly $75$ ns the qubit state loses enough coherence that the unitary correction actually worsens the overlap.

For both system temperatures the greatest benefit from correction is seen early on in the decay time, long before the cavity occupation has reached steady state. Typically one would wait for the cavity to be unoccupied before further operations on the qubit are preformed, as cavity photons are still interacting with the qubit. However, in set-ups with tunable coupling between the cavity and the qubit \cite{Wenner:2014gf,Chen:2014mz,Zeytinoglu:2015rz}, it would be possible to turn off the interaction between the cavity and the qubit once enough measurement data has been accumulated and then correct the final state of the qubit. In this way one would could achieve both more accurate and faster initialization of the qubit state via measurement and unitary correction. A similar initialization scheme involving both cavity and qubit control has recently been implemented \cite{Geerlings:2013rz}.

The analytic expressions of equations (\ref{eqn:Ug}) and (\ref{eqn:Ue}) for the correction unitaries correctly calculate the amplitude of the rotation, described by the angle $\theta$. Unfortunately, due to higher order nonlinear effects in the full Hamiltonian the analytic phases $\Sigma_g$ and $\Sigma_e$ do not give good results. To solve this problem, we performed a brute force optimization over the phase of the rotation to obtain the excellent results shown in FIGs.~\ref{fig:Cold} and \ref{fig:Hot}.

In conclusion, we have shown that during the most commonly used dispersive readout protocol for superconducting qubits a coherent rotation error is applied to the qubit, and the measurement scheme is not QND for any finite measurement time. This coherent rotation causes errors in repeated measurements and in qubit initialization; however, as we have shown, it can be corrected for by unitary feedback. This correction is most advantageous early on in the decay time, and in experiments with tunable qubit-cavity coupling our scheme shows promising results for faster and more accurate qubit initialization.

The authors acknowledge insightful discussions with Bruno G. Taketani, Daniel Sank, Karl-Peter Marzlin and John M. Martinis. Supported by the Army Research Office under contract W911NF-14-1-0080 and the European Union through ScaleQIT. LCGG acknowledges support from NSERC through an NSERC PGS-D. 

\clearpage

\appendix

\section{First Order Approximation of the Dressed Coherent States}
\label{app:DressApp}

For the ground qubit dressed coherent state, we have
\begin{align}
\nonumber\overline{\ket{g,\al}} &= e^{-\frac{\abs{\alpha}^2}{2}}\sum_{n}\frac{\alpha^n}{\sqrt{n!}}\overline{\ket{g,n}} \\ \nonumber &= e^{-\frac{\abs{\alpha}^2}{2}}\sum_{n}\frac{\alpha^n}{\sqrt{n!}}\Big( \cos\left({\lambda\sqrt{n}}\right)\ket{g,n} \\ \nonumber &- \sin\left({\lambda\sqrt{n}}\right)\ket{e,n-1}\Big) \\ 
\nonumber&= e^{-\frac{\abs{\alpha_{g}}^2}{2}}\sum_{n}\frac{\alpha^n}{\sqrt{n!}}\Bigg(\left(1-\frac{n\lambda^2}{2}\right) \ket{g,n} \\  &- \lambda\sqrt{n}\ket{e,n-1}\Bigg) + \mathcal{O}(\lambda^3).
\end{align}
Tracing out the cavity we obtain
\begin{align}
\nonumber&{\rm Tr_C}\left[\overline{\ket{g,\al}}\overline{\bra{g,\al}}\right] = \left(1-\lambda^2\abs{\al}^2\right)\ket{g}\bra{g} + \lambda^2\abs{\al}^2\ket{e}\bra{e} \\ \nonumber&- \lambda\al^*\ket{g}\bra{e}  - \lambda\al\ket{e}\bra{g} + \mathcal{O}(\lambda^3)\\
&= \left(\ket{g}-\lambda\al\ket{e}\right)\left(\bra{g}-\lambda\al^*\bra{e}\right)  -\lambda^2\abs{\al}^2\ket{g}\bra{g} + \mathcal{O}(\lambda^3).
\end{align}
We now approximate this minimally mixed state by a pure state to obtain
\be
\overline{\ket{g,\al}}  = \left(\ket{g}-\lambda\al\ket{e}\right)/\sqrt{\mathcal{N}} + \mathcal{O}(\lambda^2),
\ee
where $\mathcal{N} = 1 + \lambda^2\abs{\al}^2$ is the normalization.

Now for the excited qubit dressed coherent state, we begin with
\begin{align}
\nonumber\overline{\ket{e,\al}} &= e^{-\frac{\abs{\alpha}^2}{2}}\sum_{n}\frac{\alpha^n}{\sqrt{n!}}\overline{\ket{e,n}} \\ 
\nonumber&= e^{-\frac{\abs{\alpha}^2}{2}}\sum_{n}\frac{\alpha^n}{\sqrt{n!}}\Big( \cos\left({\lambda\sqrt{n+1}}\right)\ket{e,n}\\ 
\nonumber&+ \sin\left({\lambda\sqrt{n+1}}\right)\ket{g,n+1}\Big) \\ 
\nonumber&= e^{-\frac{\abs{\alpha}^2}{2}}\sum_{n}\frac{\alpha^n}{\sqrt{n!}}\Bigg(\left(1-\frac{(n+1)\lambda^2}{2}\right) \ket{e,n} \\ 
&+ \lambda\sqrt{n+1}\ket{g,n+1}\Bigg) + \mathcal{O}(\lambda^3).
\end{align}
Tracing out the cavity we obtain
\begin{align}
\nonumber&{\rm Tr_C}\left[\overline{\ket{e,\al}}\overline{\bra{e,\al}}\right] = \left(1-\lambda^2(1+\abs{\al}^2)\right)\ket{e}\bra{e} \\ 
\nonumber&+\lambda^2( 1+\abs{\al}^2)\ket{g}\bra{g} + \lambda\al^*\ket{g}\bra{e}  + \lambda\al\ket{e}\bra{g} + \mathcal{O}(\lambda^3)\\
\nonumber&= \left(\ket{e}+\lambda\al^*\ket{g}\right)\left(\bra{e}+\lambda\al\bra{g}\right)  \\ 
&+\lambda^2\ket{g}\bra{g}-\lambda^2(1+\abs{\al}^2)\ket{e}\bra{e}  + \mathcal{O}(\lambda^3).
\end{align}
Making the same approximation as for the previous case, we arrive at the pure state
\be
\overline{\ket{e,\al}}  = \left(\ket{e}+\lambda\al^*\ket{g}\right)/\sqrt{\mathcal{N}} + \mathcal{O}(\lambda^2),
\ee
where, as before, $\mathcal{N}$ is the normalization.

\section{Numerical Master Equation}
\label{app:NME}

To derive the master equations of equations (\ref{eqn:MEnoP}) and (\ref{eqn:ME}) we begin by considering the system-bath Hamiltonian
\be
\hat{H}_{\rm E} = \hat{H}_{\rm T}(t) + \sum_{k}\eta_k\hat{b}_k^\dagger\hat{b}_k + \sum_{k}g_k\left(\hat{a}+\hat{a}^\dagger\right)\left(\hat{b}_k+\hat{b}_k^\dagger\right)
\label{eqn:HamSB}
\ee
where the second term in equation (\ref{eqn:HamSB}) is the bath self-Hamiltonian, and the third term is the system-bath coupling. To derive an effective master equation for the system, it is appropriate to work in the instantaneous eigenbasis of $\hat{H}_{\rm T}(t)$; however, to simply things we will derive the master equation in the eigenbasis of $\hat{H}$ of equation (\ref{eqn:JC}). The difference between these two is a frame transformation by a time dependent cavity displacement, which for the parameter regime under consideration is inconsequential to the applicability of the master equation obtained.

Following the procedure of \cite{Beaudoin11} we derive an effective evolution equation for the system (ignoring the coherent evolution for the time being)
\begin{align}
&\nonumber\dot{\rho}(t) =\\ \nonumber& \sum_{\substack{j,n \\k>j\\m>n}}C_{jk}C_{nm}^*\Big(\ket{j}\bra{k}\rho(t)\ket{m}\bra{n}-\ket{m}\bra{n}\ket{j}\bra{k}\rho(t)\Big)\\ \nonumber& \times e^{i(\Delta_{jk}- \Delta_{nm})t}\int_{0}^{\infty}ds\left<\hat{b}(s)\hat{b}^{\dagger}(0)\right>e^{-i\Delta_{jk}t} \\
&+ \nonumber\sum_{\substack{j,n \\k>j\\m>n}}C_{jk}C_{nm}^*\Big(\ket{j}\bra{k}\rho(t)\ket{m}\bra{n}-\rho(t)\ket{m}\bra{n}\ket{j}\bra{k}\Big)\\ \nonumber& \times e^{i(\Delta_{jk}- \Delta_{nm})t}\int_{0}^{\infty}ds\left<\hat{b}(0)\hat{b}^{\dagger}(s)\right>e^{i\Delta_{nm}t} \\
&+
\nonumber\sum_{\substack{j,n \\k>j\\m>n}}C_{jk}^*C_{nm}\Big(\ket{k}\bra{j}\rho(t)\ket{n}\bra{m}-\ket{n}\bra{m}\ket{k}\bra{j}\rho(t)\Big)\\ \nonumber& \times e^{-i(\Delta_{jk}- \Delta_{nm})t}\int_{0}^{\infty}ds\left<\hat{b}^{\dagger}(s)\hat{b}(0)\right>e^{i\Delta_{jk}t} \\
&\nonumber+\sum_{\substack{j,n \\k>j\\m>n}}C_{jk}^*C_{nm}\Big(\ket{k}\bra{j}\rho(t)\ket{n}\bra{m}-\rho(t)\ket{n}\bra{m}\ket{k}\bra{j}\Big)\\ & \times e^{-i(\Delta_{jk}- \Delta_{nm})t}\int_{0}^{\infty}ds\left<\hat{b}^{\dagger}(0)\hat{b}(s)\right>e^{-i\Delta_{nm}t}
\label{eqn:EffEq}
\end{align}
where $\left\{\ket{j}\right\}$ is the eigenbasis of $\hat{H}$ ordered in increasing eigenenergy and given to first order in $\lambda$ by equations (\ref{eqn:aGeig}) and (\ref{eqn:aEeig}), $\Delta_{jk}$ is the frequency difference between the $j$'th and $k$'th eigenstate (Bohr frequency), $\hat{b}(s) = \sum_{k}g_k\hat{b}_ke^{-i\eta_kt}$ is the time dependent bath lowering operator, and $C_{jk} = \bra{j}\left(\hat{a}+\hat{a}^\dagger\right)\ket{k}$. We have also assumed that the bath state is a stationary state of the bath self-Hamiltonian.

Unlike in \cite{Beaudoin11}, it is not possible to make a rotating wave approximation, as in the parameter regime under consideration the eigenspectrum of $\hat{H}$ has many nearly degenerate transitions. Instead, it is possible to derive a Lindblad form master equation in a way similar to that done in the singular coupling limit \cite{Breuer:2006uq} by assuming that the nearly degenerate transitions are actually degenerate. We notice that the coefficients $C_{jk}$ are nonzero if $\ket{j} = \overline{\ket{g/e,n}}$ and $\ket{k} = \overline{\ket{g/e,n\pm 1}}$, or if $\ket{j} = \overline{\ket{g/e,n}}$ and $\ket{k} = \overline{\ket{e/g,n}}$, while all other $C_{jk}$ are zero. The former case is what we have been calling cavity decay, while the latter case is Purcell decay, and  for each decay type the energy difference $\Delta_{jk}$ between adjacent states is approximately constant ($\ww_{\rm c}$ in the former case and $\ww_{\rm q}$ in the latter). 

Therefore, we can split the sums of equation (\ref{eqn:EffEq}) into two parts, and make a secular approximation to neglect the fast oscillating cross terms between decay types to arrive at the equation
\begin{flalign}
&\nonumber\dot{\rho}(t) = \\ \nonumber& \sum^{\mathcal{C}}_{\substack{j,n \\k>j\\m>n}}C_{jk}C_{nm}^*\mathcal{D}\Big[\ket{j}\bra{k},\ket{m}\bra{n}\Big]\rho(t)\\ \nonumber& \times\Big(1+ n_{\rm th}(\ww_{\rm c},{\rm T})\Big)J(\ww_{\rm c})\ \\
&\nonumber+
\sum^{\mathcal{C}}_{\substack{j,n \\k>j\\m>n}}C_{jk}^*C_{nm}\mathcal{D}\Big[\ket{k}\bra{j},\ket{n}\bra{m}\Big]\rho(t) n_{\rm th}(\ww_{\rm c},{\rm T})J(\ww_{\rm c}) \\
&\nonumber+
\sum^{\mathcal{P}}_{\substack{j,n \\k>j\\m>n}}C_{jk}C_{nm}^* \mathcal{D}\Big[\ket{j}\bra{k},\ket{m}\bra{n}\Big]\rho(t)\\ \nonumber& \times \Big(1+ n_{\rm th}(\ww_{\rm q},{\rm T})\Big)J(\ww_{\rm q}) \\
&+
\sum^{\mathcal{P}}_{\substack{j,n \\k>j\\m>n}}C_{jk}^*C_{nm}\mathcal{D}\Big[\ket{k}\bra{j},\ket{n}\bra{m}\Big]\rho(t) n_{\rm th}(\ww_{\rm q},{\rm T})J(\ww_{\rm q}),
\label{eqn:EffEq2}
\end{flalign}
where we have ignored the Lamb shifts, and the superscripts $\mathcal{C}$ and $\mathcal{P}$ indicate summation over cavity decay transitions and over Purcell decay transitions respectively. For more compact notation, we have also defined the two operator ``dissipator''
\be
\mathcal{D}\left[\hat{O}_1,\hat{O}_2\right]\rho(t) = \hat{O}_1\rho(t)\hat{O}_2-\frac{1}{2}\Big\{\hat{O}_2\hat{O}_1,\rho(t)\Big\}.
\ee
Following the usual procedure of Fermi's golden rule we have made the identification
\begin{align}
&\nonumber\int_{0}^{\infty}ds\left(\left<\hat{b}(s)\hat{b}^{\dagger}(0)\right>e^{-i\ww t} + \left<\hat{b}(0)\hat{b}^{\dagger}(s)\right>e^{i\ww t}\right) \\ &= \Big(1+ n_{\rm th}(\ww,{\rm T})\Big)J(\ww), \\
&\nonumber\int_{0}^{\infty}ds\left(\left<\hat{b}^{\dagger}(s)\hat{b}(0)\right>e^{i\ww t} +\left<\hat{b}^{\dagger}(0)\hat{b}(s)\right>e^{-i\ww t}\right) \\ &= n_{\rm th}(\ww,{\rm T})J(\ww),
\end{align}
where $J(\ww)$ is the spectral density of the bath, and $n_{\rm th}(\ww,{\rm T})$ is the Bose function evaluated at frequency $\ww$ and temperature T.

If we choose a global Ohmic spectral density $J(\ww) = \ww/Q_{\rm F}$, then using the relations
\begin{align}
\nonumber&\sum_{j,k>j}C_{jk}\ket{j}\bra{k} = \hat{a}, \\
&\sum_{j,k>j}C_{kj}\ket{k}\bra{j} = \sum_{j,k>j}C_{jk}^*\ket{k}\bra{j} = \hat{a}^{\dagger}.
\label{eqn:CofRels}
\end{align}
we can write equation (\ref{eqn:EffEq2}) in Lindblad form
\begin{align}
\nonumber\dot{\rho}(t) &= \Big(1+ n_{\rm th}(\ww_{\rm c},{\rm T})\Big)\left(\kappa\mathcal{D}\Big[\hat{a}_{\mathcal{C}}\Big]+\gamma_{\rm P}\mathcal{D}\Big[\hat{a}_{\mathcal{P}}\Big] \right)\rho(t) \\&+ n_{\rm th}(\ww_{\rm q},{\rm T})\left( \kappa\mathcal{D}\left[\hat{a}^{\dagger}_{\mathcal{C}}\right]+\gamma_{\rm P}\mathcal{D}\left[\hat{a}^{\dagger}_{\mathcal{P}}\right]\right)\rho(t)
\label{eqn:PurLin}
\end{align}
where $\kappa = \ww_{\rm c}/Q_{\rm F}$ is the cavity decay rate and $\gamma_{\rm P} = \lambda^2\ww_{\rm q}/Q_{\rm F}$ is the Purcell decay rate. The operators $\hat{a}_{\mathcal{C}}$ and $\hat{a}_{\mathcal{P}}$ are defined by
\begin{align}
&\hat{a}_{\mathcal{C}} = \sum_{j,k>j}^{\mathcal{C}}C_{jk}\ket{j}\bra{k} = \hat{a} - \sum_{j,k>j}^{\mathcal{P}}C_{jk}\ket{j}\bra{k}, \label{eqn:OpNoPur} \\
&\hat{a}_{\mathcal{P}} = \sum_{j,k>j}^{\mathcal{P}}C_{jk}\ket{j}\bra{k} = \hat{a} - \sum_{j,k>j}^{\mathcal{C}}C_{jk}\ket{j}\bra{k},  \label{eqn:OpPur}
\end{align}
and describe the cavity and Purcell decay processes respectively.

Equation (\ref{eqn:PurLin}) is valid for $t \ll t_{\rm ME}$, where $t_{\rm ME} = 1/\ww_{\rm max}$ describes the timescale over which $e^{i(\Delta_{jk}- \Delta_{nm})t} = e^{i\ww_{\rm max}t}$ is no longer unity, with $\ww_{\rm max}$ the largest degeneracy between transitions that were assumed to be degenerate. For cavity decay $\ww_{\rm max} \propto N\chi\lambda^2$, while for Purcell decay $\ww_{\rm max} \propto N\chi$, with $N$ the photon number of the largest occupied state. For the parameters under consideration ($\chi \leq 10$ MHz, $\lambda \leq 10^{-1}$) we have $t_{\rm ME}  \sim 10/N$ $\mu$s for cavity decay, and $t_{ME} \propto 100/N$ ns for Purcell decay. For typical experimental parameters the simulation length is on the order of 100s of nanoseconds, well below the limit imposed by cavity decay for reasonable values of $N$, but unfortunately on the same order as that set by Purcell decay. As such, we expect that the simulations of equation (\ref{eqn:PurLin}) are somewhat nonphysical; however, as the Purcell decay rate is quite small, we do expect the results to remain mostly trustworthy.

If instead of choosing a global Ohmic spectral density, we  choose a spectral density for which $J(\ww_{\rm c}) = \ww_{\rm c}/Q_{\rm F}$ but $J(\ww_{\rm q}) \approx 0$, such as has been achieved in contemporary dispersive readout schemes by a filter \cite{Reed10,Jeffrey:2014zr}, then the resulting Lindblad equation contains only cavity decay, and is given by
\begin{align}
\nonumber\dot{\rho}(t) &= \kappa\Big(1+ n_{\rm th}(\ww_{\rm c},{\rm T})\Big)\mathcal{D}\Big[\hat{a}_{\mathcal{C}}\Big]\rho(t) \\ &+ \kappa n_{\rm th}(\ww_{\rm c},{\rm T})\mathcal{D}\left[\hat{a}^{\dagger}_{\mathcal{C}}\right]\rho(t).
\label{eqn:NoPurLin}
\end{align}
This is the master equation used in the main text to described dispersive readout in a Purcell filtered system.

\section{Readout of the Undressed Qubit Excited State}
\label{app:UndressedE}

In the main text we considered starting with the excited initial state $\overline{\ket{e,0}}$, which is the first excited state in the energy eigenbasis of the qubit-cavity system. However, for some quantum information protocols it is advantageous to work in the qubit logical basis, where the logical excited state is the state $\ket{e,0}$. Such a state can be prepared by initializing the qubit-cavity system into its ground state $\ket{g,0}$, followed by a short non-adiabatic qubit pulse that flips the state of the qubit.

If we now attempt to read out the state of the qubit, from Ref.~\cite{longpaper} we see that starting in the state $\ket{e,0}$, after an applied cavity drive of the form of equation (\ref{eqn:CavDri}) the state of the system is
\begin{align}
\nonumber&\ket{\Psi(t_{\rm d})} = \cos\left(\lambda\right)\overline{\ket{e,\al_{\rm e}(t_{\rm d})}} \\ &- e^{iG(t_{\rm d})}\sin\left(\lambda\right)\hat{U}^{\dagger}_{\rm D}\ket{g}e^{-i\left(\ww_{\rm c}\hat{a}^{\dagger}\hat{a} -\chi\hat{\sigma}_z \hat{a}^{\dagger}\hat{a}\right)t_{\rm d}}\hat{D}(\al'_{\rm g}(t_{\rm d}))\ket{1},
\label{eqn:undressedEfinal}
\end{align}
where $\al'_{\rm g}(t_{\rm d}) = \eps^*\left(e^{-i\chi t_{\rm d}}-1\right)/\chi$, $\hat{D}(\beta)$ is the usual displacement operator for a harmonic oscillator, $\hat{U}_{\rm D} = \exp\left\{\lambda\left(\hat{\sigma}^{+}\hat{a}-\hat{\sigma}^{-}\hat{a}^{\dagger}\right)\right\}$ is the dispersive frame transformation operator, and $e^{iG(t_{\rm d})}$ is a relative qubit phase whose form is unimportant \cite{longpaper}. From equation (\ref{eqn:undressedEfinal}) we see that the final state contains a component for which the qubit is in its ground state, and the frequency of the cavity signal for this component will be close to $\ww_g$, measurement of which indicates the qubit is in its ground state. This introduces the possibility of misidentifying the qubit state; however, the amplitude of the ground state component is small as it as scales with $\sin^2(\lambda) \approx \lambda^2$. Nevertheless, this sets a fundamental limit for the readout fidelity of the undressed excited state via standard dispersive readout as presented here, and partly explains the unequal readout fidelities reported in \cite{Jeffrey:2014zr}.

After sufficiently long measurement of the frequency $\ww_e$, the state of equation (\ref{eqn:undressedEfinal}) will have collapsed to the dressed coherent state $\overline{\ket{e,\al_{\rm e}(t_{\rm d})}}$, and the rest of the readout and coherent feedback protocol can occur as described in the main text, without further modification.\\

\section{Purcell Decay}
\label{app:Pdecay}
\begin{figure}[h]
\includegraphics[width = \columnwidth]{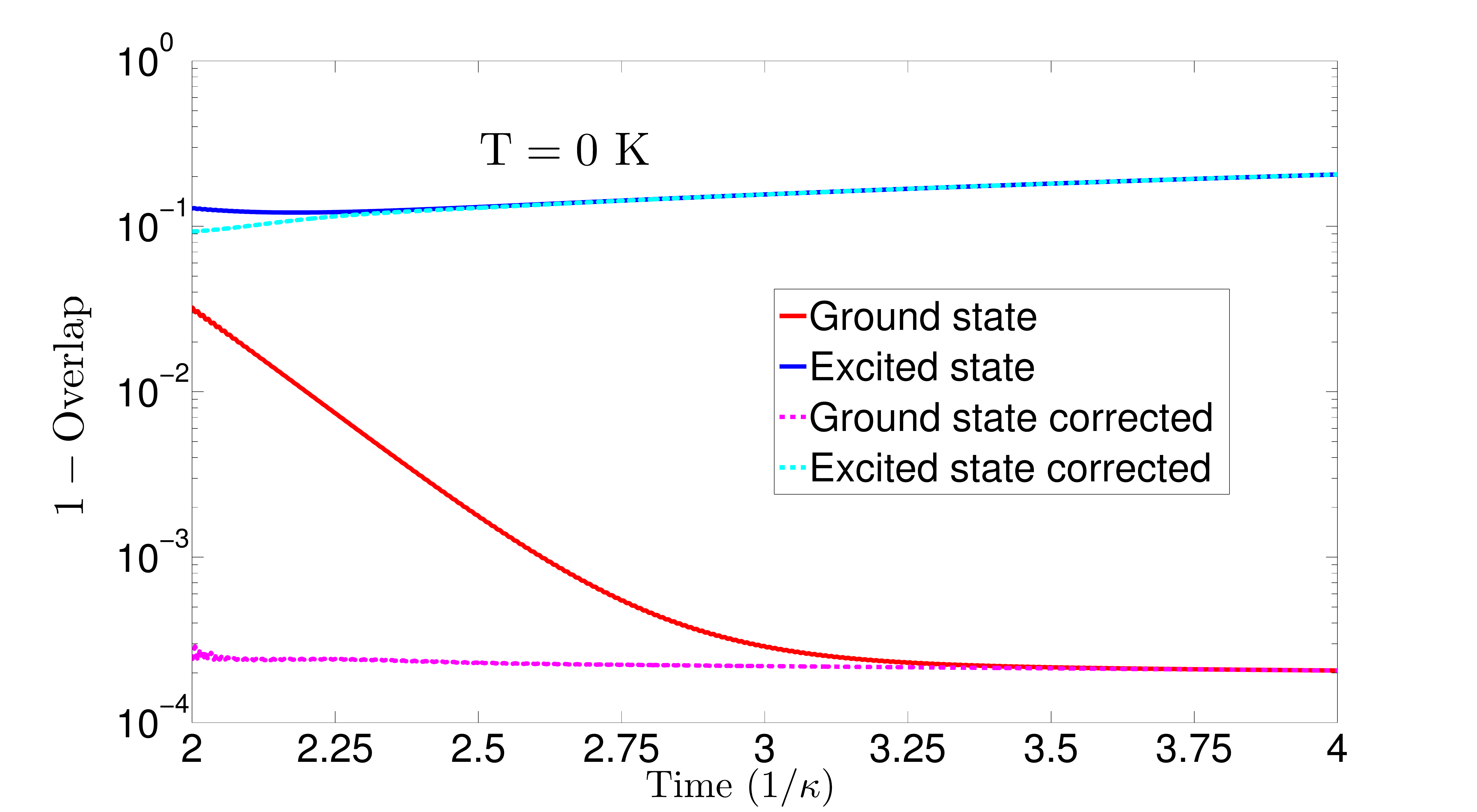}
\caption{$1 - \mathcal{F}_{\nu}(\tau)$ and $1 - \mathcal{F}_{\nu}^{\rm C}(\tau)$ for T = 0. A drive strength of $\abs{\eps}/2\pi = 0.04$ GHz, a cavity decay rate of $1/\kappa = 100$ ns, and $\abs{\lambda} = 0.1$ were used for this simulation.}
\label{fig:ColdPur}
\end{figure}

To examine the effect of Purcell decay we simulate the unfiltered master equation
\begin{align}
\nonumber&\dot{\rho}(t) = -i\left[\hat{H}_{\rm T}(t),\rho(t)\right] \\ \nonumber&+ \kappa\left(\left(1+ n_{\rm th}(\ww_{\rm c},{\rm T})\right)\mathcal{D}\Big[\hat{a}_{\mathcal{C}}\Big] + n_{\rm th}(\ww_{\rm c},{\rm T})\mathcal{D}\left[\hat{a}^{\dagger}_{\mathcal{C}}\right]\right)\rho(t) \\&+ \gamma_{\rm P}\left(\left(1+ n_{\rm th}(\ww_{\rm q},{\rm T})\right)\mathcal{D}\Big[\hat{a}_{\mathcal{P}}\Big] + n_{\rm th}(\ww_{\rm q},{\rm T})\mathcal{D}\left[\hat{a}^{\dagger}_{\mathcal{P}}\right]\right)\rho(t),
\label{eqn:ME}
\end{align}
where the operator $\hat{a}_{\mathcal{P}}$ of equation (\ref{eqn:OpPur}) describes Purcell decay. As before, we simulate the evolution for the initial states $\ket{g,0}$ and $\overline{\ket{e,0}}$, at a temperature T = 0. The overlap error for the uncorrected and corrected states are shown in FIG.~\ref{fig:ColdPur}. As can be seen, for the initial state $\ket{g,0}$ the results remain unchanged, as Purcell decay only minimally affects the decay of states of the form $\overline{\ket{g,\al}}$. Coherent feedback can still be used in this case to correct the qubit state, with excellent improvement in the overlap. For $\overline{\ket{e,0}}$ the results are quite different from that of FIG.~\ref{fig:Cold}, as Purcell decay drives the system to the global ground state $\ket{g,0}$ and, as such the qubit state decays. In this case the coherent correction does little good, as the qubit state is lost by Purcell decay.

\bibliography{PhaseEffects}

\begin{thebibliography}{31}
\expandafter\ifx\csname natexlab\endcsname\relax\def\natexlab#1{#1}\fi
\expandafter\ifx\csname bibnamefont\endcsname\relax
  \def\bibnamefont#1{#1}\fi
\expandafter\ifx\csname bibfnamefont\endcsname\relax
  \def\bibfnamefont#1{#1}\fi
\expandafter\ifx\csname citenamefont\endcsname\relax
  \def\citenamefont#1{#1}\fi
\expandafter\ifx\csname url\endcsname\relax
  \def\url#1{\texttt{#1}}\fi
\expandafter\ifx\csname urlprefix\endcsname\relax\def\urlprefix{URL }\fi
\providecommand{\bibinfo}[2]{#2}
\providecommand{\eprint}[2][]{\url{#2}}

\bibitem[{\citenamefont{Braginsky and Khalili}(1992)}]{Braginsky:1992kq}
\bibinfo{author}{\bibfnamefont{V.~B.} \bibnamefont{Braginsky}}
  \bibnamefont{and} \bibinfo{author}{\bibfnamefont{F.~Y.}
  \bibnamefont{Khalili}}, \emph{\bibinfo{title}{Quantum Measurement}}
  (\bibinfo{publisher}{Cambridge University Press}, \bibinfo{address}{New York,
  NY, USA}, \bibinfo{year}{1992}), ISBN \bibinfo{isbn}{0-521-41928-X}.

\bibitem[{\citenamefont{Blais et~al.}(2004)\citenamefont{Blais, Huang,
  Wallraff, Girvin, and Schoelkopf}}]{Blais2004}
\bibinfo{author}{\bibfnamefont{A.}~\bibnamefont{Blais}},
  \bibinfo{author}{\bibfnamefont{R.-S.} \bibnamefont{Huang}},
  \bibinfo{author}{\bibfnamefont{A.}~\bibnamefont{Wallraff}},
  \bibinfo{author}{\bibfnamefont{S.~M.} \bibnamefont{Girvin}},
  \bibnamefont{and} \bibinfo{author}{\bibfnamefont{R.~J.}
  \bibnamefont{Schoelkopf}}, \bibinfo{journal}{Phys. Rev. A}
  \textbf{\bibinfo{volume}{69}}, \bibinfo{pages}{062320}
  (\bibinfo{year}{2004}).

\bibitem[{\citenamefont{Johansson et~al.}(2006)\citenamefont{Johansson,
  Tornberg, and Wilson}}]{Johansson:2006vn}
\bibinfo{author}{\bibfnamefont{G.}~\bibnamefont{Johansson}},
  \bibinfo{author}{\bibfnamefont{L.}~\bibnamefont{Tornberg}}, \bibnamefont{and}
  \bibinfo{author}{\bibfnamefont{C.~M.} \bibnamefont{Wilson}},
  \bibinfo{journal}{Phys. Rev. B} \textbf{\bibinfo{volume}{74}},
  \bibinfo{pages}{100504} (\bibinfo{year}{2006}).

\bibitem[{\citenamefont{Siddiqi et~al.}(2004)\citenamefont{Siddiqi, Vijay,
  Pierre, Wilson, Metcalfe, Rigetti, Frunzio, and Devoret}}]{Siddiqi:2004fj}
\bibinfo{author}{\bibfnamefont{I.}~\bibnamefont{Siddiqi}},
  \bibinfo{author}{\bibfnamefont{R.}~\bibnamefont{Vijay}},
  \bibinfo{author}{\bibfnamefont{F.}~\bibnamefont{Pierre}},
  \bibinfo{author}{\bibfnamefont{C.~M.} \bibnamefont{Wilson}},
  \bibinfo{author}{\bibfnamefont{M.}~\bibnamefont{Metcalfe}},
  \bibinfo{author}{\bibfnamefont{C.}~\bibnamefont{Rigetti}},
  \bibinfo{author}{\bibfnamefont{L.}~\bibnamefont{Frunzio}}, \bibnamefont{and}
  \bibinfo{author}{\bibfnamefont{M.~H.} \bibnamefont{Devoret}},
  \bibinfo{journal}{Phys. Rev. Lett.} \textbf{\bibinfo{volume}{93}},
  \bibinfo{pages}{207002} (\bibinfo{year}{2004}).

\bibitem[{\citenamefont{Castellanos-Beltran and Lehnert}(2007)}]{Castellanos07}
\bibinfo{author}{\bibfnamefont{M.~A.} \bibnamefont{Castellanos-Beltran}}
  \bibnamefont{and} \bibinfo{author}{\bibfnamefont{K.~W.}
  \bibnamefont{Lehnert}}, \bibinfo{journal}{Appl. Phys. Lett.}
  \textbf{\bibinfo{volume}{91}}, \bibinfo{pages}{083509}
  (\bibinfo{year}{2007}).

\bibitem[{\citenamefont{Bergeal et~al.}(2010)\citenamefont{Bergeal, Schackert,
  Metcalfe, Vijay, Manucharyan, Frunzio, Prober, Schoelkopf, Girvin, and
  Devoret}}]{Bergeal:2010ty}
\bibinfo{author}{\bibfnamefont{N.}~\bibnamefont{Bergeal}},
  \bibinfo{author}{\bibfnamefont{F.}~\bibnamefont{Schackert}},
  \bibinfo{author}{\bibfnamefont{M.}~\bibnamefont{Metcalfe}},
  \bibinfo{author}{\bibfnamefont{R.}~\bibnamefont{Vijay}},
  \bibinfo{author}{\bibfnamefont{V.~E.} \bibnamefont{Manucharyan}},
  \bibinfo{author}{\bibfnamefont{L.}~\bibnamefont{Frunzio}},
  \bibinfo{author}{\bibfnamefont{D.~E.} \bibnamefont{Prober}},
  \bibinfo{author}{\bibfnamefont{R.~J.} \bibnamefont{Schoelkopf}},
  \bibinfo{author}{\bibfnamefont{S.~M.} \bibnamefont{Girvin}},
  \bibnamefont{and} \bibinfo{author}{\bibfnamefont{M.~H.}
  \bibnamefont{Devoret}}, \bibinfo{journal}{Nature}
  \textbf{\bibinfo{volume}{465}}, \bibinfo{pages}{64} (\bibinfo{year}{2010}).

\bibitem[{\citenamefont{Bishop et~al.}(2010)\citenamefont{Bishop, Ginossar, and
  Girvin}}]{Bishop:2010qf}
\bibinfo{author}{\bibfnamefont{L.~S.} \bibnamefont{Bishop}},
  \bibinfo{author}{\bibfnamefont{E.}~\bibnamefont{Ginossar}}, \bibnamefont{and}
  \bibinfo{author}{\bibfnamefont{S.~M.} \bibnamefont{Girvin}},
  \bibinfo{journal}{Phys. Rev. Lett.} \textbf{\bibinfo{volume}{105}},
  \bibinfo{pages}{100505} (\bibinfo{year}{2010}).

\bibitem[{\citenamefont{Reed et~al.}(2010{\natexlab{a}})\citenamefont{Reed,
  DiCarlo, Johnson, Sun, Schuster, Frunzio, and Schoelkopf}}]{Reed:2010kq}
\bibinfo{author}{\bibfnamefont{M.~D.} \bibnamefont{Reed}},
  \bibinfo{author}{\bibfnamefont{L.}~\bibnamefont{DiCarlo}},
  \bibinfo{author}{\bibfnamefont{B.~R.} \bibnamefont{Johnson}},
  \bibinfo{author}{\bibfnamefont{L.}~\bibnamefont{Sun}},
  \bibinfo{author}{\bibfnamefont{D.~I.} \bibnamefont{Schuster}},
  \bibinfo{author}{\bibfnamefont{L.}~\bibnamefont{Frunzio}}, \bibnamefont{and}
  \bibinfo{author}{\bibfnamefont{R.~J.} \bibnamefont{Schoelkopf}},
  \bibinfo{journal}{Phys. Rev. Lett.} \textbf{\bibinfo{volume}{105}},
  \bibinfo{pages}{173601} (\bibinfo{year}{2010}{\natexlab{a}}).

\bibitem[{\citenamefont{Mutus et~al.}(2014)\citenamefont{Mutus, White, Barends,
  Chen, Chen, Chiaro, Dunsworth, Jeffrey, Kelly, Megrant
  et~al.}}]{Mutus:2014yg}
\bibinfo{author}{\bibfnamefont{J.~Y.} \bibnamefont{Mutus}},
  \bibinfo{author}{\bibfnamefont{T.~C.} \bibnamefont{White}},
  \bibinfo{author}{\bibfnamefont{R.}~\bibnamefont{Barends}},
  \bibinfo{author}{\bibfnamefont{Y.}~\bibnamefont{Chen}},
  \bibinfo{author}{\bibfnamefont{Z.}~\bibnamefont{Chen}},
  \bibinfo{author}{\bibfnamefont{B.}~\bibnamefont{Chiaro}},
  \bibinfo{author}{\bibfnamefont{A.}~\bibnamefont{Dunsworth}},
  \bibinfo{author}{\bibfnamefont{E.}~\bibnamefont{Jeffrey}},
  \bibinfo{author}{\bibfnamefont{J.}~\bibnamefont{Kelly}},
  \bibinfo{author}{\bibfnamefont{A.}~\bibnamefont{Megrant}},
  \bibnamefont{et~al.}, \bibinfo{journal}{Applied Physics Letters}
  \textbf{\bibinfo{volume}{104}}, \bibinfo{eid}{263513} (\bibinfo{year}{2014}).

\bibitem[{\citenamefont{O'Brien et~al.}(2014)\citenamefont{O'Brien, Macklin,
  Siddiqi, and Zhang}}]{OBrien:2014yg}
\bibinfo{author}{\bibfnamefont{K.}~\bibnamefont{O'Brien}},
  \bibinfo{author}{\bibfnamefont{C.}~\bibnamefont{Macklin}},
  \bibinfo{author}{\bibfnamefont{I.}~\bibnamefont{Siddiqi}}, \bibnamefont{and}
  \bibinfo{author}{\bibfnamefont{X.}~\bibnamefont{Zhang}},
  \bibinfo{journal}{Phys. Rev. Lett.} \textbf{\bibinfo{volume}{113}},
  \bibinfo{pages}{157001} (\bibinfo{year}{2014}).

\bibitem[{\citenamefont{White et~al.}(2015)\citenamefont{White, Mutus, Hoi,
  Barends, Campbell, Chen, Chen, Chiaro, Dunsworth, Jeffrey
  et~al.}}]{White2015}
\bibinfo{author}{\bibfnamefont{T.~C.} \bibnamefont{White}},
  \bibinfo{author}{\bibfnamefont{J.~Y.} \bibnamefont{Mutus}},
  \bibinfo{author}{\bibfnamefont{I.-C.} \bibnamefont{Hoi}},
  \bibinfo{author}{\bibfnamefont{R.}~\bibnamefont{Barends}},
  \bibinfo{author}{\bibfnamefont{B.}~\bibnamefont{Campbell}},
  \bibinfo{author}{\bibfnamefont{Y.}~\bibnamefont{Chen}},
  \bibinfo{author}{\bibfnamefont{Z.}~\bibnamefont{Chen}},
  \bibinfo{author}{\bibfnamefont{B.}~\bibnamefont{Chiaro}},
  \bibinfo{author}{\bibfnamefont{A.}~\bibnamefont{Dunsworth}},
  \bibinfo{author}{\bibfnamefont{E.}~\bibnamefont{Jeffrey}},
  \bibnamefont{et~al.}, \bibinfo{journal}{Applied Physics Letters}
  \textbf{\bibinfo{volume}{106}}, \bibinfo{eid}{242601} (\bibinfo{year}{2015}).

\bibitem[{\citenamefont{Vijay et~al.}(2011)\citenamefont{Vijay, Slichter, and
  Siddiqi}}]{Vijay2011}
\bibinfo{author}{\bibfnamefont{R.}~\bibnamefont{Vijay}},
  \bibinfo{author}{\bibfnamefont{D.~H.} \bibnamefont{Slichter}},
  \bibnamefont{and} \bibinfo{author}{\bibfnamefont{I.}~\bibnamefont{Siddiqi}},
  \bibinfo{journal}{Phys. Rev. Lett.} \textbf{\bibinfo{volume}{106}},
  \bibinfo{pages}{110502} (\bibinfo{year}{2011}).

\bibitem[{\citenamefont{Johnson et~al.}(2012)\citenamefont{Johnson, Macklin,
  Slichter, Vijay, Weingarten, Clarke, and Siddiqi}}]{Johnson12}
\bibinfo{author}{\bibfnamefont{J.~E.} \bibnamefont{Johnson}},
  \bibinfo{author}{\bibfnamefont{C.}~\bibnamefont{Macklin}},
  \bibinfo{author}{\bibfnamefont{D.~H.} \bibnamefont{Slichter}},
  \bibinfo{author}{\bibfnamefont{R.}~\bibnamefont{Vijay}},
  \bibinfo{author}{\bibfnamefont{E.~B.} \bibnamefont{Weingarten}},
  \bibinfo{author}{\bibfnamefont{J.}~\bibnamefont{Clarke}}, \bibnamefont{and}
  \bibinfo{author}{\bibfnamefont{I.}~\bibnamefont{Siddiqi}},
  \bibinfo{journal}{Phys. Rev. Lett.} \textbf{\bibinfo{volume}{109}},
  \bibinfo{pages}{050506} (\bibinfo{year}{2012}).

\bibitem[{\citenamefont{Rist\`e et~al.}(2012)\citenamefont{Rist\`e, van
  Leeuwen, Ku, Lehnert, and DiCarlo}}]{Riste12}
\bibinfo{author}{\bibfnamefont{D.}~\bibnamefont{Rist\`e}},
  \bibinfo{author}{\bibfnamefont{J.~G.} \bibnamefont{van Leeuwen}},
  \bibinfo{author}{\bibfnamefont{H.-S.} \bibnamefont{Ku}},
  \bibinfo{author}{\bibfnamefont{K.~W.} \bibnamefont{Lehnert}},
  \bibnamefont{and} \bibinfo{author}{\bibfnamefont{L.}~\bibnamefont{DiCarlo}},
  \bibinfo{journal}{Phys. Rev. Lett.} \textbf{\bibinfo{volume}{109}},
  \bibinfo{pages}{050507} (\bibinfo{year}{2012}).

\bibitem[{\citenamefont{Rist\`{e} et~al.}(2013)\citenamefont{Rist\`{e},
  Dukalski, Watson, de~Lange, Tiggelman, Blanter, Lehnert, Schouten, and
  DiCarlo}}]{Riste2013}
\bibinfo{author}{\bibfnamefont{D.}~\bibnamefont{Rist\`{e}}},
  \bibinfo{author}{\bibfnamefont{M.}~\bibnamefont{Dukalski}},
  \bibinfo{author}{\bibfnamefont{C.~A.} \bibnamefont{Watson}},
  \bibinfo{author}{\bibfnamefont{G.}~\bibnamefont{de~Lange}},
  \bibinfo{author}{\bibfnamefont{M.~J.} \bibnamefont{Tiggelman}},
  \bibinfo{author}{\bibfnamefont{Y.~M.} \bibnamefont{Blanter}},
  \bibinfo{author}{\bibfnamefont{K.~W.} \bibnamefont{Lehnert}},
  \bibinfo{author}{\bibfnamefont{R.~N.} \bibnamefont{Schouten}},
  \bibnamefont{and} \bibinfo{author}{\bibfnamefont{L.}~\bibnamefont{DiCarlo}},
  \bibinfo{journal}{Nature} \textbf{\bibinfo{volume}{502}},
  \bibinfo{pages}{350} (\bibinfo{year}{2013}).

\bibitem[{\citenamefont{Chow et~al.}(2014)\citenamefont{Chow, Gambetta,
  Magesan, Abraham, Cross, Johnson, Masluk, Ryan, Smolin, Srinivasan
  et~al.}}]{Chow:2014fk}
\bibinfo{author}{\bibfnamefont{J.~M.} \bibnamefont{Chow}},
  \bibinfo{author}{\bibfnamefont{J.~M.} \bibnamefont{Gambetta}},
  \bibinfo{author}{\bibfnamefont{E.}~\bibnamefont{Magesan}},
  \bibinfo{author}{\bibfnamefont{D.~W.} \bibnamefont{Abraham}},
  \bibinfo{author}{\bibfnamefont{A.~W.} \bibnamefont{Cross}},
  \bibinfo{author}{\bibfnamefont{B.~R.} \bibnamefont{Johnson}},
  \bibinfo{author}{\bibfnamefont{N.~A.} \bibnamefont{Masluk}},
  \bibinfo{author}{\bibfnamefont{C.~A.} \bibnamefont{Ryan}},
  \bibinfo{author}{\bibfnamefont{J.~A.} \bibnamefont{Smolin}},
  \bibinfo{author}{\bibfnamefont{S.~J.} \bibnamefont{Srinivasan}},
  \bibnamefont{et~al.}, \bibinfo{journal}{Nat Commun}
  \textbf{\bibinfo{volume}{5}}, \bibinfo{pages}{4015} (\bibinfo{year}{2014}).

\bibitem[{\citenamefont{Steffen et~al.}(2013)\citenamefont{Steffen, Salathe,
  Oppliger, Kurpiers, Baur, Lang, Eichler, Puebla-Hellmann, Fedorov, and
  Wallraff}}]{Steffen2013}
\bibinfo{author}{\bibfnamefont{L.}~\bibnamefont{Steffen}},
  \bibinfo{author}{\bibfnamefont{Y.}~\bibnamefont{Salathe}},
  \bibinfo{author}{\bibfnamefont{M.}~\bibnamefont{Oppliger}},
  \bibinfo{author}{\bibfnamefont{P.}~\bibnamefont{Kurpiers}},
  \bibinfo{author}{\bibfnamefont{M.}~\bibnamefont{Baur}},
  \bibinfo{author}{\bibfnamefont{C.}~\bibnamefont{Lang}},
  \bibinfo{author}{\bibfnamefont{C.}~\bibnamefont{Eichler}},
  \bibinfo{author}{\bibfnamefont{G.}~\bibnamefont{Puebla-Hellmann}},
  \bibinfo{author}{\bibfnamefont{A.}~\bibnamefont{Fedorov}}, \bibnamefont{and}
  \bibinfo{author}{\bibfnamefont{A.}~\bibnamefont{Wallraff}},
  \bibinfo{journal}{Nature} \textbf{\bibinfo{volume}{500}},
  \bibinfo{pages}{319} (\bibinfo{year}{2013}).

\bibitem[{\citenamefont{Jeffrey et~al.}(2014)\citenamefont{Jeffrey, Sank,
  Mutus, White, Kelly, Barends, Chen, Chen, Chiaro, Dunsworth
  et~al.}}]{Jeffrey:2014zr}
\bibinfo{author}{\bibfnamefont{E.}~\bibnamefont{Jeffrey}},
  \bibinfo{author}{\bibfnamefont{D.}~\bibnamefont{Sank}},
  \bibinfo{author}{\bibfnamefont{J.~Y.} \bibnamefont{Mutus}},
  \bibinfo{author}{\bibfnamefont{T.~C.} \bibnamefont{White}},
  \bibinfo{author}{\bibfnamefont{J.}~\bibnamefont{Kelly}},
  \bibinfo{author}{\bibfnamefont{R.}~\bibnamefont{Barends}},
  \bibinfo{author}{\bibfnamefont{Y.}~\bibnamefont{Chen}},
  \bibinfo{author}{\bibfnamefont{Z.}~\bibnamefont{Chen}},
  \bibinfo{author}{\bibfnamefont{B.}~\bibnamefont{Chiaro}},
  \bibinfo{author}{\bibfnamefont{A.}~\bibnamefont{Dunsworth}},
  \bibnamefont{et~al.}, \bibinfo{journal}{Phys. Rev. Lett.}
  \textbf{\bibinfo{volume}{112}}, \bibinfo{pages}{190504}
  (\bibinfo{year}{2014}).

\bibitem[{\citenamefont{Govia and Wilhelm}(2015)}]{longpaper}
\bibinfo{author}{\bibfnamefont{L.~C.~G.} \bibnamefont{Govia}} \bibnamefont{and}
  \bibinfo{author}{\bibfnamefont{F.~K.} \bibnamefont{Wilhelm}}
  (\bibinfo{year}{2015}), \bibinfo{note}{arXiv:1506.04997}.

\bibitem[{\citenamefont{Hutchison et~al.}(2009)\citenamefont{Hutchison,
  Gambetta, Blais, and Wilhelm}}]{Hutchison:2009qy}
\bibinfo{author}{\bibfnamefont{C.~L.} \bibnamefont{Hutchison}},
  \bibinfo{author}{\bibfnamefont{J.~M.} \bibnamefont{Gambetta}},
  \bibinfo{author}{\bibfnamefont{A.}~\bibnamefont{Blais}}, \bibnamefont{and}
  \bibinfo{author}{\bibfnamefont{F.~K.} \bibnamefont{Wilhelm}},
  \bibinfo{journal}{Canadian Journal of Physics} \textbf{\bibinfo{volume}{87}},
  \bibinfo{pages}{225} (\bibinfo{year}{2009}).

\bibitem[{\citenamefont{Jaynes and Cummings}(1963)}]{Jaynes63}
\bibinfo{author}{\bibfnamefont{E.}~\bibnamefont{Jaynes}} \bibnamefont{and}
  \bibinfo{author}{\bibfnamefont{F.}~\bibnamefont{Cummings}},
  \bibinfo{journal}{IEEE Proc.} \textbf{\bibinfo{volume}{51}},
  \bibinfo{pages}{89} (\bibinfo{year}{1963}).

\bibitem[{\citenamefont{Haroche and Raimond}(2006)}]{Haroche06}
\bibinfo{author}{\bibfnamefont{S.}~\bibnamefont{Haroche}} \bibnamefont{and}
  \bibinfo{author}{\bibfnamefont{J.-M.} \bibnamefont{Raimond}},
  \emph{\bibinfo{title}{Exploring the Quantum: Atoms, Cavities, and Photons}}
  (\bibinfo{publisher}{Oxford University Press}, \bibinfo{address}{Oxford},
  \bibinfo{year}{2006}).

\bibitem[{\citenamefont{Beaudoin et~al.}(2011)\citenamefont{Beaudoin, Gambetta,
  and Blais}}]{Beaudoin11}
\bibinfo{author}{\bibfnamefont{F.}~\bibnamefont{Beaudoin}},
  \bibinfo{author}{\bibfnamefont{J.~M.} \bibnamefont{Gambetta}},
  \bibnamefont{and} \bibinfo{author}{\bibfnamefont{A.}~\bibnamefont{Blais}},
  \bibinfo{journal}{Phys. Rev. A} \textbf{\bibinfo{volume}{84}},
  \bibinfo{pages}{043832} (\bibinfo{year}{2011}).

\bibitem[{\citenamefont{Sete et~al.}(2014)\citenamefont{Sete, Gambetta, and
  Korotkov}}]{Sete:2014fk}
\bibinfo{author}{\bibfnamefont{E.~A.} \bibnamefont{Sete}},
  \bibinfo{author}{\bibfnamefont{J.~M.} \bibnamefont{Gambetta}},
  \bibnamefont{and} \bibinfo{author}{\bibfnamefont{A.~N.}
  \bibnamefont{Korotkov}}, \bibinfo{journal}{Phys. Rev. B}
  \textbf{\bibinfo{volume}{89}}, \bibinfo{pages}{104516}
  (\bibinfo{year}{2014}).

\bibitem[{\citenamefont{Reed et~al.}(2010{\natexlab{b}})\citenamefont{Reed,
  Johnson, Houck, DiCarlo, Chow, Schuster, Frunzio, and Schoelkopf}}]{Reed10}
\bibinfo{author}{\bibfnamefont{M.}~\bibnamefont{Reed}},
  \bibinfo{author}{\bibfnamefont{B.}~\bibnamefont{Johnson}},
  \bibinfo{author}{\bibfnamefont{A.}~\bibnamefont{Houck}},
  \bibinfo{author}{\bibfnamefont{L.}~\bibnamefont{DiCarlo}},
  \bibinfo{author}{\bibfnamefont{J.}~\bibnamefont{Chow}},
  \bibinfo{author}{\bibfnamefont{D.}~\bibnamefont{Schuster}},
  \bibinfo{author}{\bibfnamefont{L.}~\bibnamefont{Frunzio}}, \bibnamefont{and}
  \bibinfo{author}{\bibfnamefont{R.}~\bibnamefont{Schoelkopf}},
  \bibinfo{journal}{Appl. Phys. Lett.} \textbf{\bibinfo{volume}{96}},
  \bibinfo{pages}{203110} (\bibinfo{year}{2010}{\natexlab{b}}).

\bibitem[{\citenamefont{Fowler et~al.}(2012)\citenamefont{Fowler, Mariantoni,
  Martinis, and Cleland}}]{Fowler:2012sf}
\bibinfo{author}{\bibfnamefont{A.~G.} \bibnamefont{Fowler}},
  \bibinfo{author}{\bibfnamefont{M.}~\bibnamefont{Mariantoni}},
  \bibinfo{author}{\bibfnamefont{J.~M.} \bibnamefont{Martinis}},
  \bibnamefont{and} \bibinfo{author}{\bibfnamefont{A.~N.}
  \bibnamefont{Cleland}}, \bibinfo{journal}{Phys. Rev. A}
  \textbf{\bibinfo{volume}{86}}, \bibinfo{pages}{032324}
  (\bibinfo{year}{2012}).

\bibitem[{\citenamefont{Wenner et~al.}(2014)\citenamefont{Wenner, Yin, Chen,
  Barends, Chiaro, Jeffrey, Kelly, Megrant, Mutus, Neill
  et~al.}}]{Wenner:2014gf}
\bibinfo{author}{\bibfnamefont{J.}~\bibnamefont{Wenner}},
  \bibinfo{author}{\bibfnamefont{Y.}~\bibnamefont{Yin}},
  \bibinfo{author}{\bibfnamefont{Y.}~\bibnamefont{Chen}},
  \bibinfo{author}{\bibfnamefont{R.}~\bibnamefont{Barends}},
  \bibinfo{author}{\bibfnamefont{B.}~\bibnamefont{Chiaro}},
  \bibinfo{author}{\bibfnamefont{E.}~\bibnamefont{Jeffrey}},
  \bibinfo{author}{\bibfnamefont{J.}~\bibnamefont{Kelly}},
  \bibinfo{author}{\bibfnamefont{A.}~\bibnamefont{Megrant}},
  \bibinfo{author}{\bibfnamefont{J.~Y.} \bibnamefont{Mutus}},
  \bibinfo{author}{\bibfnamefont{C.}~\bibnamefont{Neill}},
  \bibnamefont{et~al.}, \bibinfo{journal}{Phys. Rev. Lett.}
  \textbf{\bibinfo{volume}{112}}, \bibinfo{pages}{210501}
  (\bibinfo{year}{2014}).

\bibitem[{\citenamefont{Chen et~al.}(2014)\citenamefont{Chen, Neill, Roushan,
  Leung, Fang, Barends, Kelly, Campbell, Chen, Chiaro et~al.}}]{Chen:2014mz}
\bibinfo{author}{\bibfnamefont{Y.}~\bibnamefont{Chen}},
  \bibinfo{author}{\bibfnamefont{C.}~\bibnamefont{Neill}},
  \bibinfo{author}{\bibfnamefont{P.}~\bibnamefont{Roushan}},
  \bibinfo{author}{\bibfnamefont{N.}~\bibnamefont{Leung}},
  \bibinfo{author}{\bibfnamefont{M.}~\bibnamefont{Fang}},
  \bibinfo{author}{\bibfnamefont{R.}~\bibnamefont{Barends}},
  \bibinfo{author}{\bibfnamefont{J.}~\bibnamefont{Kelly}},
  \bibinfo{author}{\bibfnamefont{B.}~\bibnamefont{Campbell}},
  \bibinfo{author}{\bibfnamefont{Z.}~\bibnamefont{Chen}},
  \bibinfo{author}{\bibfnamefont{B.}~\bibnamefont{Chiaro}},
  \bibnamefont{et~al.}, \bibinfo{journal}{Phys. Rev. Lett.}
  \textbf{\bibinfo{volume}{113}}, \bibinfo{pages}{220502}
  (\bibinfo{year}{2014}).

\bibitem[{\citenamefont{Zeytino\ifmmode~\breve{g}\else \u{g}\fi{}lu
  et~al.}(2015)\citenamefont{Zeytino\ifmmode~\breve{g}\else \u{g}\fi{}lu,
  Pechal, Berger, Abdumalikov, Wallraff, and Filipp}}]{Zeytinoglu:2015rz}
\bibinfo{author}{\bibfnamefont{S.}~\bibnamefont{Zeytino\ifmmode~\breve{g}\else
  \u{g}\fi{}lu}}, \bibinfo{author}{\bibfnamefont{M.}~\bibnamefont{Pechal}},
  \bibinfo{author}{\bibfnamefont{S.}~\bibnamefont{Berger}},
  \bibinfo{author}{\bibfnamefont{A.~A.} \bibnamefont{Abdumalikov}},
  \bibinfo{author}{\bibfnamefont{A.}~\bibnamefont{Wallraff}}, \bibnamefont{and}
  \bibinfo{author}{\bibfnamefont{S.}~\bibnamefont{Filipp}},
  \bibinfo{journal}{Phys. Rev. A} \textbf{\bibinfo{volume}{91}},
  \bibinfo{pages}{043846} (\bibinfo{year}{2015}).

\bibitem[{\citenamefont{Geerlings et~al.}(2013)\citenamefont{Geerlings,
  Leghtas, Pop, Shankar, Frunzio, Schoelkopf, Mirrahimi, and
  Devoret}}]{Geerlings:2013rz}
\bibinfo{author}{\bibfnamefont{K.}~\bibnamefont{Geerlings}},
  \bibinfo{author}{\bibfnamefont{Z.}~\bibnamefont{Leghtas}},
  \bibinfo{author}{\bibfnamefont{I.~M.} \bibnamefont{Pop}},
  \bibinfo{author}{\bibfnamefont{S.}~\bibnamefont{Shankar}},
  \bibinfo{author}{\bibfnamefont{L.}~\bibnamefont{Frunzio}},
  \bibinfo{author}{\bibfnamefont{R.~J.} \bibnamefont{Schoelkopf}},
  \bibinfo{author}{\bibfnamefont{M.}~\bibnamefont{Mirrahimi}},
  \bibnamefont{and} \bibinfo{author}{\bibfnamefont{M.~H.}
  \bibnamefont{Devoret}}, \bibinfo{journal}{Phys. Rev. Lett.}
  \textbf{\bibinfo{volume}{110}}, \bibinfo{pages}{120501}
  (\bibinfo{year}{2013}).

\bibitem[{\citenamefont{Breuer and Petruccione}(2006)}]{Breuer:2006uq}
\bibinfo{author}{\bibfnamefont{H.-P.} \bibnamefont{Breuer}} \bibnamefont{and}
  \bibinfo{author}{\bibfnamefont{F.}~\bibnamefont{Petruccione}},
  \emph{\bibinfo{title}{The Theory of Open Quantum Systems}}
  (\bibinfo{publisher}{Oxford University Press}, \bibinfo{year}{2006}).

\end{thebibliography}

\end{document}